\documentclass[a4paper,11pt]{article}

\usepackage[utf8]{inputenc}   
\usepackage{amsmath}          
\usepackage{amssymb}          
\usepackage{graphicx}         
\usepackage{color}
\usepackage{hyperref}         
\usepackage{geometry}         
\usepackage{natbib}           
\usepackage{authblk}          
\usepackage{appendix}
\usepackage{multirow}

\title{Be X-ray binary pulsar SXP 138: non-linear spin-down, pulse and spectral characteristics}
\author[1]{Soham Pravin Sanyashiv}
\author[2]{Sayantan Bhattacharya\thanks{
sayantan34@gmail.com}}
\author[2]{Sudip Bhattacharyya}
\affil[1]{Indian Institute of Science Education and Research Kolkata, Mohanpur, West Bengal, 741246, India}
\affil[2]{Department of Astronomy and Astrophysics, Tata Institute of Fundamental Research, 1 Homi Bhabha Road, Colaba, Mumbai 400005, India}
\date{\today}

\begin{document}

\maketitle

\begin{abstract}
We study the timing and spectral properties of the Be/X-ray binary pulsar SXP 138 using four \textit{NuSTAR} observations spanning 2016 to 2017. Analysis of the light curves using the Lomb-Scargle periodogram shows an increase in the spin period of SXP 138 from 140.69 to 140.85 seconds,
indicating that the source is in the propeller regime. We calculate the associated rate of spin period change and characterize its non-linearity with a quadratic fit. Pulse profiles obtained by folding the light curves at the spin period show two primary high peaks and two secondary peaks for all the observations. Such features in the pulse profile can result from the combined effect of pencil and fan beam emissions from the two antipodal hotspots accompanied by the relativistic bending of photons. The energy spectra fit with both the blackbody and the power-law spectral model. The best-fit models of all observations show an overall increasing trend in the temperature of the source from 1.8 keV to 2.5 keV, during the first three observations, and a possible decreasing trend in the photon index from 1.8 to 1.7. The power-law component is dominant in the later observations, which is associated with an increase in Compton scattering and the accretion rate. SXP 138 has not been studied in detail, and this work bridges the gap by providing the first comprehensive timing and spectral analysis of this source. These findings improve our understanding of its accretion processes and emission mechanisms, placing SXP 138 in the broader context of Be/X-ray binaries.
\end{abstract}

\twocolumn
\section{Introduction}
X-ray binaries are systems consisting of a compact object orbiting around a companion star while accreting its matter. Depending on the mass of the donor star, the X-ray binaries can be classified as high-mass X-ray binaries (HMXBs) and low-mass X-ray binaries (LMXBs). HMXBs contain O-type or B-type mass donors with masses $>10 M_\odot$ \citep{liu2006catalogue}, whereas LXMB has mass donors with a spectral type after A, with a mass of a few solar masses or less \citep{liu2007catalogue}. The type of mass donors determines the mass transfer process. Low-mass companions, including a few HXMBs, are usually associated with mass transfers via Roche lobe overflow, whereas most HXMBs lose mass via line-driven winds.

The Be/X-ray binaries (BeXRBs), representing the largest sub-class of HXMBs, are composed of a compact object (most often a neutron star (NS)) and a Be star as an optical companion with an eccentric and wide orbit \citep{reig2011x}. The optical and infrared (IR) emission originating from the equatorial circumstellar disk around a rapidly rotating Be star is characterized by Balmer series emission lines and IR excess \citep{porter2003classical,reig2011x}. When the compact object approaches the mass donor, enhanced accretion takes place and the accreting material interacts with the NS accretion disk, resulting in X-ray outbursts. We observe two types of X-ray outbursts from BeXRBs. With a maximum luminosity of $\leq 10^{37}$ erg $s^{-1}$, the type I outbursts are short and quasi-periodic, occurring close to the periastron passage of the binary system. Type II are irregular outbursts and reach peak luminosities of $\geq 10^{37} - 10^{39}$ erg $s^{-1}$ and persist for multiple or significant portions of the orbit.

Multiple surveys reveal that a significant number of BeXRBs are present in the Milky Way and the Magellanic Clouds, with $\sim$81 BeXRBs hosted by our galaxy and $\sim$70 known in the Small Magellanic Cloud (SMC) \citep{liu2006catalogue};\citep{reig2011x};\citep{coe2015catalogue}. Although remarkably lower in mass compared to our galaxy, SMC provides a large sample size. It has a well-constrained distance and intergalactic absorption, making it an excellent laboratory to study both the components in the binary systems. SXP 138 is such a BeXRB system located in SMC (RA: 00h 53m 23.86s, DEC: -72d 27m 15.1s). Initial observations of SXP 138 using \textit{Chandra} revealed a spin period of $138.04 \pm 0.61$ s and an orbital period of $125\pm 1.5$ days around the optical companion Be star [MA93]667 \citep{edge2004three}. Continuous RXTE observations of SXP 138 also displayed a superorbital period of about 1000 days with a possibility of stochastic changes in the accretion disk as a cause for this periodicity \citep{chashkina2015superorbital}. In recent studies, it is observed that SXP 138 is one of such sources that accretes in the Rayleigh-Taylor stable regime for all observations \citep{o2025observing}.

The accretion properties and geometry of the pulsar and its hotspot(s) can be efficiently studied using pulse profiles. Each source shows unique behavior as well as variations in its pulse profiles during an outburst, which indicates a change in the emission geometry of the accretion columns. This results in dramatic changes in the shape and properties of their pulse profiles. Since multiple non-linear processes are involved in studying the properties of a neutron star and its accretion column, providing a complete description of the pulse profiles is extremely complex. An attempt to classify the X-ray pulsars based on their observed pulse profiles was performed by \citet{alonso2022common}. Further insights into the physics of the accretion-powered sources can be obtained through spectral analysis. Generally, the spectrum of accretion-powered sources is modeled to contain multi-component functions, which include blackbody, power-law components, and a high-energy exponential cutoff. For a detailed discussion on the model for the spectral formation process in HMXBs, refer to \citet{becker2007thermal}.

The existing literature on SXP 138 is very limited, and this work aims to bridge the gap and characterize this interesting source based on its spectral and timing properties. In this paper, we study the evolution of spin period, pulsar beaming geometry, and spectral properties of SXP 138 by analyzing four \textit{NuSTAR} observations from 2016 to 2017. Section~\ref{Methods} outlines the details of the observations and data analysis methods. The results of timing and spectral analysis are reported in Section~\ref{3}. The discussion and conclusion are presented in Section~\ref{4} and Section~\ref{5}, respectively.

\begin{figure*}[t]
\centering
\includegraphics[scale=0.35]{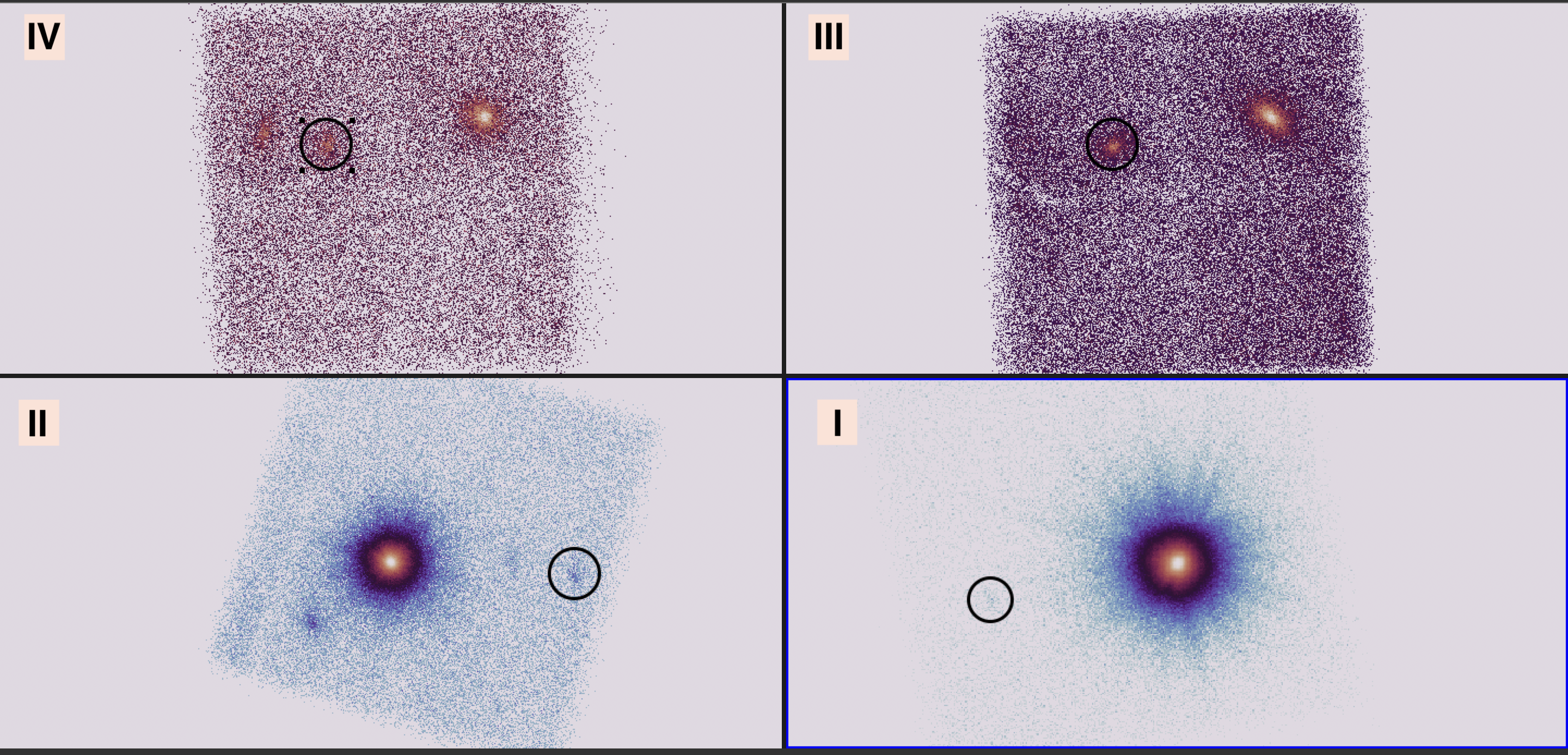}
\caption{\textit{NuSTAR}-FPMA image of all observations used, in the energy range 3-50 keV. The location of the source, SXP 138, is indicated by a circular region.\label{figure1} (see section~\ref{Methods} for detailed methods)}
\label{fig:sub2}
\end{figure*}

\begin{table*}[t]
\caption{Log of all the observations of SXP 138 with \textit{NuSTAR} spanning 2016-2017.\label{table1}}
\begin{tabular}{cllcll}
\hline
\multirow{2}{*}{Observation Number} & \multirow{2}{*}{Observation ID} & \multirow{2}{*}{Date (Time)} & \multirow{2}{*}{Exposure (s)} & \multicolumn{2}{c}{Counts} \\ \cline{5-6} 
 &  &  &  & \multicolumn{1}{l}{FPMA} & FPMB \\ \hline \hline
I & 90201035002 & 2016-08-13 (18:26:12) & 38602 & \multicolumn{1}{l}{1709} & 1562 \\
II & 30361001002 & 2017-04-12 (18:46:13) & 69565 & \multicolumn{1}{l}{3868} & 2672 \\
III & 50311003002 & 2017-05-03 (02:51:13) & 172390 & \multicolumn{1}{l}{25879} & 19545 \\
IV & 50311003004 & 2017-08-07 (02:41:08) & 87065 & \multicolumn{1}{l}{5176} & 3909 \\ \hline
\end{tabular}
\end{table*}

\section{Methodology}\label{Methods}

We analyze \textit{NuSTAR} observations of SXP 138 in this study. The \textit{NuSTAR} mission was launched on 12th June 2012, becoming the first focusing hard X-ray astronomical observatory (\cite{harrison2013nuclear}). The central component of the \textit{NuSTAR} instrument is composed of two co-aligned grazing-incidence hard X-ray telescopes, FPMA and FPMB. The instrument is designed to operate in the energy band from 3 to 79 keV. To achieve significant reflectance of the soft to hard X-ray photons and optimize the broadband energy response, the mirrors in each optic are coated with depth-graded multilayers of Pt/C and W/Si, with a CdZnTe pixel detector present at the focal point of each telescope. The observatory has a temporal resolution of $2\mu s$, while its energy resolution is 400 eV at 10 keV and 900 eV at 68 keV.

Using Heasoft 6.33 software, we perform data reduction with standard analysis pipelines. The nupipeline task is used to process unfiltered events in the presence of the CALDB of version 20240826. We separately obtain the source and background data products for FPMA and FPMB from the cleaned event files using the 'nuproducts' software modules. We add the background-subtracted light curves for both focal-plane modules for further analysis. We follow the above procedure to analyze \textit{NuSTAR} observations of SXP 138. In this paper, we retrieve four observations from the public HEASARC archive spanning 2016-2017, numbered I, II, III, and IV (Table~\ref{table1}). Figure~\ref{figure1} shows the source image for all observations. For each observation, we obtain light curves for the soft energy band (3-8 keV), hard energy band (8-50 keV), and the broad energy band (3-50 keV).\\

We extract and fit the energy spectrum of SXP 138 in the 3-50 keV range using XPSEC 12.14\footnote{ Xspec: An X-Ray Spectral Fitting Package developed by: Keith Arnaud, Ben Dorman, and Craig Gordon HEASARC Software Development, Astrophysics Science Divsion\citep{arnaud1999xspec}}. The energy channels are discarded depending on the instrument-specific flags using the \textit{ignore bad} command. All the obtained spectra are grouped so that each bin contains at least 30 counts. We use the following models to fit the energy spectra: blackbody (bb), power law photon spectrum (po), cut-off power law (cutoffpl), and Tuebingen-Boulder ISM absorption model (tbabs). We perform simultaneous fitting for the energy spectra of FPMA and FPMB.

\begin{figure*}[t]
\centering
\includegraphics[scale=0.8]{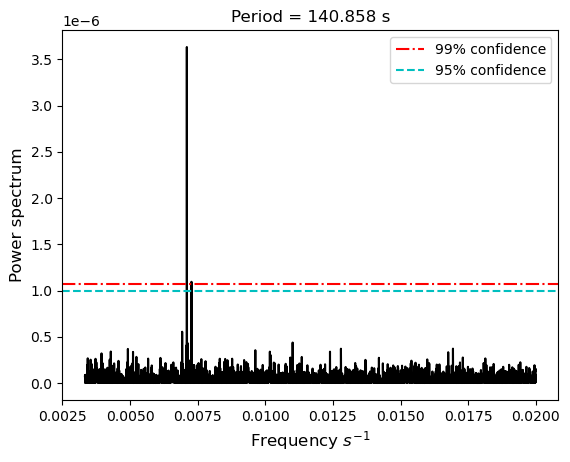}
\caption{Lomb-Scargle Periodogram of broadband light curve of \textit{NuSTAR} Obs IV of SXP 138. The horizontal red dash-dotted line gives the power where the False Alarm Probability is 0.01. Similarly, the cyan dashed line gives the power where the False Alarm Probability is 0.05. We observe the highest peak at the frequency $\sim 7.1 \times 10^{-3}s^{-1}$(Refer to Section~\ref{3} for detail).\label{figure2}}
\label{fig:sub2}
\end{figure*}

\section{RESULTS\label{3}}
\subsection{TIMING ANALYSIS}\label{3.1}
We obtain the spin periods and the pulse profiles of SXP 138 using standard Fourier transform methods to search for X-ray pulsations in the 3-60 keV \textit{NuSTAR} light curves. The spin period of SXP 138 is detected in the soft, hard, and broad energy bands for each observation using the Lomb-Scargle (LS) periodogram algorithm (\cite{vanderplas2018understanding}). The significance of the acquired spin period is quantified using the False Alarm Probability (FAP) of the normalized Lomb-Scargle periodogram(\cite{horne1986prescription}). 

\begin{figure*}[t]
\centering
\begin{minipage}{0.49\textwidth}
    \centering
    \includegraphics[width=0.8\textwidth]{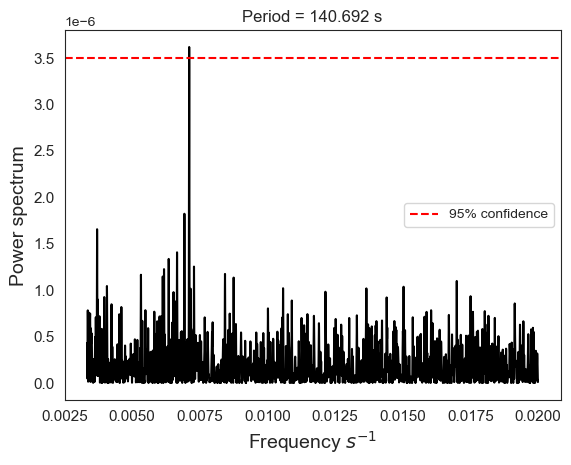}
\end{minipage}
\begin{minipage}{0.49\textwidth}
    \centering
    \includegraphics[width=1\textwidth]{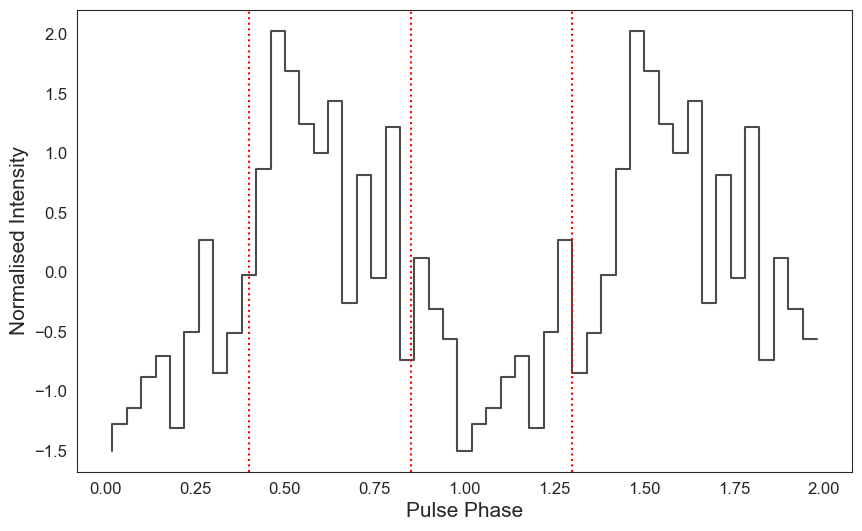}
\end{minipage}
\caption{\textit{Left:} Periodogram of the {NuSTAR} light-curve of SXP 138  obtained for Obs I. The horizontal red dash-dotted line gives the power where the False Alarm Probability is 0.05 indicating the highest and the most significant peak. \textit{Right:} Pulse profile obtained from \textit{NuSTAR} Obs I of SXP 138 in a broadband energy spectrum. The red dotted vertical lines divide the pulse phase domain into regions with two high primary peaks with positive normalized intensity (phase 0.4-0.85) and regions with two positive-valued secondary peaks, as well as one negative-valued central peak (phase 0.85-1.3).  (Refer to Section~\ref{3.1}). \label{figure3}}
\label{fig:test}
\end{figure*}

Standard error estimation methods (assumption of Gaussian errors) are not directly applicable in these Fourier-based approaches, and any uncertainties are dominated by systematic effects rather than statistical noise. Hence, we present the periods only with significantly low FAP (95\% significance) and omitted error bars. An example of the periodogram for Obs IV in the broad energy spectrum is shown in Figure~\ref{figure2}. The periodogram shows that the highest peak is the most significant peak, with a FAP less than 0.01, giving a spin period of 140.868 seconds. Obs I shows a significant peak only in the broad-band spectrum, giving a spin period of 140.692 seconds (Figure~\ref{figure3}). For other observations, we average spin periods of the energy bands that gives significant period detections. The average pulse period of SXP 138 is estimated to be 140.707s, 140.729s, and 140.858s for Obs II, Obs III, and Obs IV, respectively.  

We see a clear increase in the spin period of SXP 138. We define the time of Obs I as the initial time (t=0). By fitting a linear equation to describe the evolution of the spin period ($P_{\rm spin}$), we obtain:
\begin{align}
P_{\rm spin} = q_1t+q_2
\end{align} 
where $q_1 = 4.27\times 10^{-9} ss^{-1}$ and $q_2 = 140.67 s$, with a Root Mean Square Error of 0.0623.

The nonlinear evolution leads us to use a quadratic model. The spin periods follow a quadratic fit with the parameters:
\begin{align}
P_{\rm spin} = p_1t^2+p_2t + p_3
\end{align} where $p_1 = 4.56\times 10^{-16}s^{-1}$, $p_2 = -8.79\times10^{-9}ss^{-1}$, and $p_3 = 140.69s$, with Root Mean Square Error of 0.0018. Hence, this quadratic fit performs better than the linear fit.

\begin{figure}[t]
\centering
\includegraphics[scale=0.5]{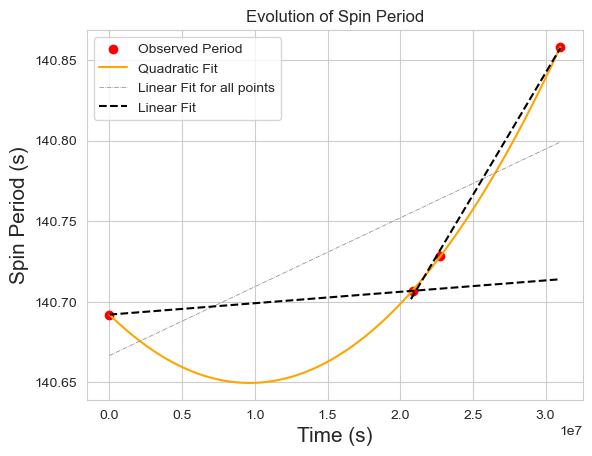}
\caption{The diagram shows the evolution of the spin period of SXP 138 with linear (grey dashed-dotted) and quadratic fit (orange solid line). The two linear fits (black dashed lines) show the variation in the rate of change in the spin period. Here the time of Obs I is referred to as t=0. A description of the parameter values of the fits is given in Section~\ref{3.1}.\label{figure4}}
\label{fig:sub2}
\end{figure}

To analyze the variation in the rate of change in the spin period, we perform two separate linear fits, grouping the first two and last three spin period values (the second value is common for both). The former yields a fit with a slope of $7.07\times10^{-10}ss^{-1}$ and a y-intercept of $140.69s$. The latter shows a slope of $1.515\times 10^{-8}ss^{-1}$ and a y-intercept of $140.39s$ with a Root Mean Square Error of 0.0041. The increase in slope shows a rise in the spin period rate of SXP 138. The evolution of the spin period with its model fits is shown in Figure~\ref{figure4}, and the associated physics is discussed in section~\ref{evol}.

\begin{table*}[tbh]
\centering 
\caption{Average period and partial fractions of \textit{NuSTAR} Obs II, III and IV of SXP 138$^*$.\label{table2}}
\begin{tabular}{|l|l|lll|}
\hline
\multirow{2}{*}{Observation} & \multirow{2}{*}{Average Period (s)} & \multicolumn{3}{l|}{Pulse Fraction(PF)(\%)} \\ \cline{3-5} 
 &  & \multicolumn{1}{l|}{Soft Band} & \multicolumn{1}{l|}{Hard Band} & Broad Band \\ \hline
Obs II & 140.707  & \multicolumn{1}{l|}{51 $\pm$ 1} & \multicolumn{1}{l|}{30 $\pm$ 1} & 40 $\pm$ 2 \\
Obs III & 140.729  & \multicolumn{1}{l|}{44 $\pm$ 3} & \multicolumn{1}{l|}{27 $\pm$ 4} & 31 $\pm$ 5 \\
Obs IV & 140.854  & \multicolumn{1}{l|}{44 $\pm$ 1} & \multicolumn{1}{l|}{36 $\pm$ 2} & 38 $\pm$ 2 \\ \hline
\end{tabular}\newline

\footnotesize $^*$All soft Band and broad band light curves give the highest significant peak with at least 95\% confidence for all three observations. The hard energy spectrum for Obs II (significance of 1\%)  and Obs IV (significance of 85\%) showed very high FAP and hence was not used to \\ calculate the average spin period.
\end{table*}

\begin{figure*}[tbh]
\centering
\includegraphics[scale=0.4]{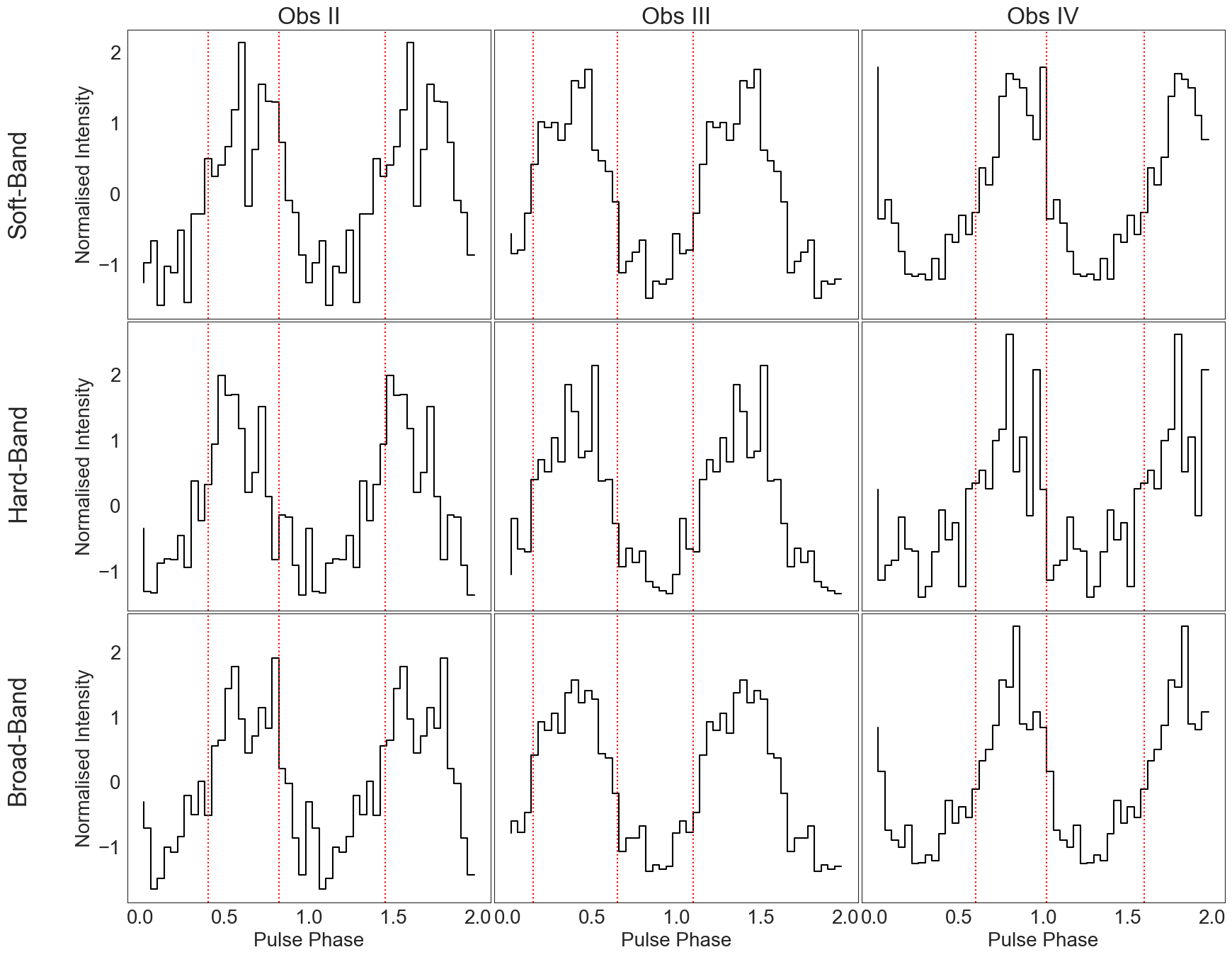}
\caption{Pulse profiles(solid black) for 2017 observations across each energy bands. For each observation, the red dotted lines divide the pulse phase domain into regions containing two high peaks (Obs II: 0.4-0.8, Obs III: 0.15-0.6, Obs IV: 0.6-1.1,) and two secondary peaks (Obs II: 0.9-1.45, Obs III: 0.65-1.1, Obs IV: 1.1-1.6). Refer to Section~\ref{3.1}  \label{figure5}}
\label{fig:sub2}
\end{figure*}

Utilizing the derived spin periods, we generate pulse profiles for the specified energy bands during the observations. Since we are primarily interested in studying the shape, we perform a normalization of the pulse profile $P(\phi)$:
\begin{align}
P_{\rm N}(\phi) = \frac{P-<P>}{\sigma}
\end{align} where $<P>$ represents the mean of the pulse profile $P$ and $\sigma$ is the standard deviation of the pulse profile. Figure~\ref{figure5} shows the pulse profile of SXP 138 for Obs II, Obs III, and Obs IV. We observe the following behavior exhibited by the pulse profiles:
\begin{itemize}
    \item[] All pulse profiles show two high primary peaks with positive normalized intensity, which are closely spaced to each other. These peaks appear in phases 0.4-0.85, 0.15-0.6, and 0.6-1.1 in Obs II, Obs III, and Obs IV, respectively. The intensity differences between the two peaks vary across the energy band. For Obs IV, the left-side peak transitions from being the lower of the two to the highest peak as we go from the soft band to the broadband spectrum. This trend is observed for the right-hand peak for Obs II. In Obs III, the right-hand peak remains the highest across energy bands. It can be argued that decreasing the bin size might smoothen the additional peaks and reveal one strong peak but that might also cause filtration of actual physical effects. We observe that the peaks are fairly consistent over different observations.  
    \item[] The pulse profiles also show two secondary peaks with a negative normalized intensity. These peaks are seen in phases 0.9-1.45, 0.65-1.1, and 1.1-1.6 in Obs II, Obs III, and Obs IV, respectively. This feature is significantly stronger in Obs II and III compared to Obs IV, where the sharp secondary peaks only appear in the hard band. 
    \item[] In Obs I, the two primary high peaks, and the two secondary peaks are present in phases 0.4-0.85 and 0.85-1.3, respectively (Figure~\ref{figure3}). However, in this case, while the secondary peaks have positive normalized intensity, a smaller central peak with a negative normalized intensity is also observed.
\end{itemize}

The pulse fraction measures the pulsated part present in the total observed emission. For each pulse profile, we obtain the Pulse Fractions (PF) by adopting the definition that calculates the relative amplitude of pulsed modulation
\begin{equation}
    \texttt{PF} = \frac{P_{\rm max}-P_{\rm min}}{P_{\rm max}+P_{\rm min}}
\end{equation} where $P_{max}$ and $P_{min}$ are the maximum and minimum intensities of the pulse profile ($P(\phi)$) respectively. The pulse fraction obtained for each observation at all energy bands is given in Table~\ref{table2}. For all the observations, the Soft band spectrum exhibits the highest pulse fraction whereas the Hard band spectrum shows the lowest pulse fraction. This indicates that the emissions from SXP 138 are predominantly composed of soft photons. 

\subsection{SPECTRAL ANALYSIS\label{3.2}}

The broadband spectra of SXP 138 are fitted using the four XSPEC models: bb*tbabs, po*tbabs, (bb+po)*tbabs, (bb+cutoffpl)*tbabs. The model with the best fit is determined by the statistical values of the fit. XSPEC calculates the reduced $\chi^2$ ($\chi^2_r$) using the ratio of $\chi^2$ and the degree of freedom (\texttt{d.o.f}), which is obtained by subtracting the number of model parameters from the number of channels. 
The model with $\chi^2_r$ closest to 1 is considered to be the best fit for the energy spectrum. We acquire null hypothesis probability(p), which gives the probability of obtaining a value of $\chi^2$ that is as large or larger than the observed value, assuming the model is accurate. The model with the highest null hypothesis probability is considered to be the best fit. As SMC is known to have low line of sight absorption \citep{tobrej2025spectro}, we set the hydrogen column, $nH$, to a fixed value of $0.5 \times 10^{22}$. All model parameters for FPMB are tied to those of FPMA as a result of the simultaneous fitting. Table~\ref{Para} and Table~\ref{Stats} provide the parameter values and statistical values to assess the fitness of the models for all the observations, respectively. Although the combination of the blackbody spectrum and cutoff power-law component also shows statistically acceptable values for a fit, its parameters are associated with unphysical values and errors. Hence, we do not consider (bb+cutoffpl)*tbabs to be a good model for any observation. \\

\begin{figure*}[h!]
\centering
\begin{minipage}{0.49\textwidth}
    \centering
    \includegraphics[width=1.01\textwidth]{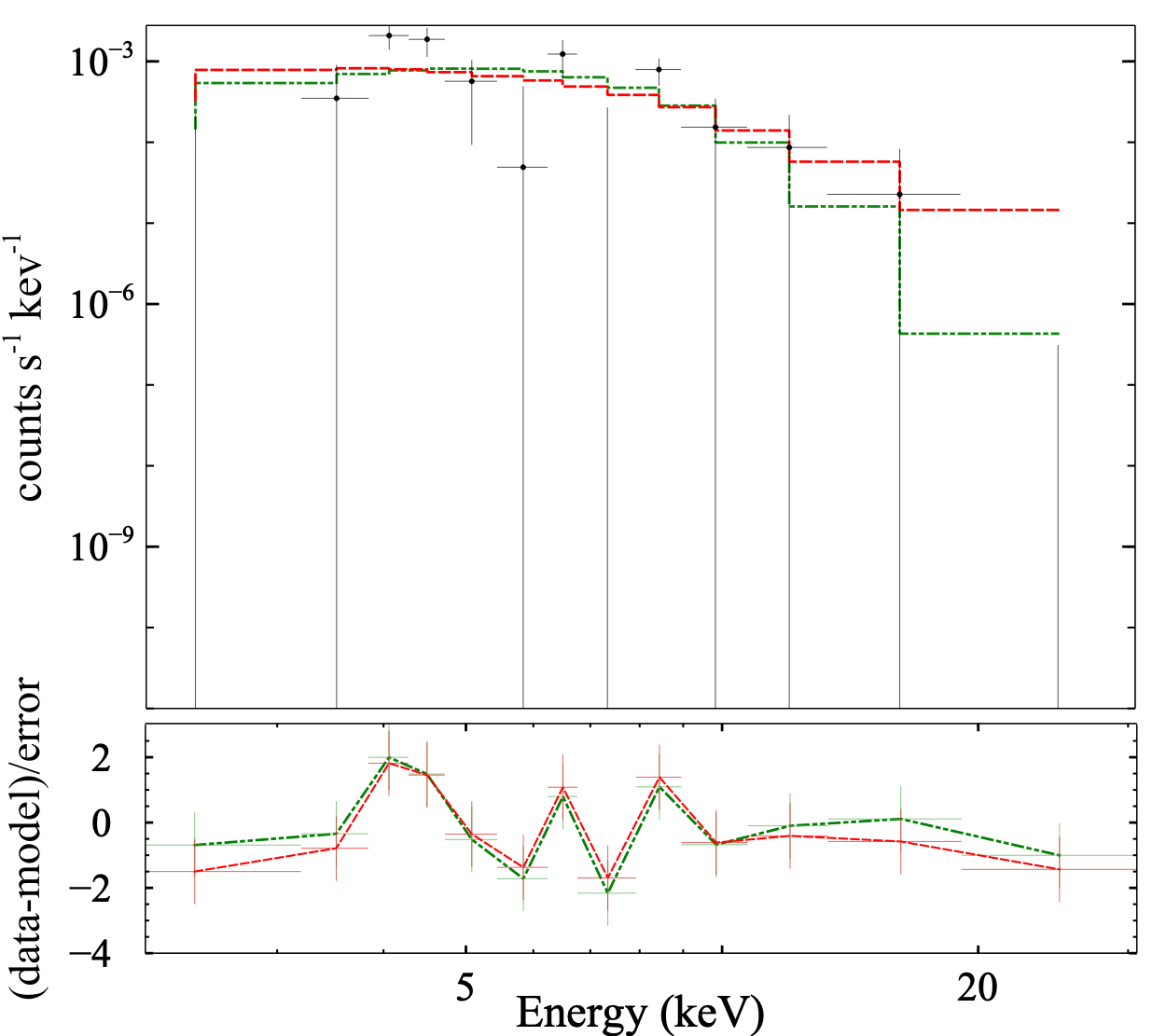}
\end{minipage}
\begin{minipage}{0.49\textwidth}
    \centering
    \includegraphics[width=1.01\textwidth]{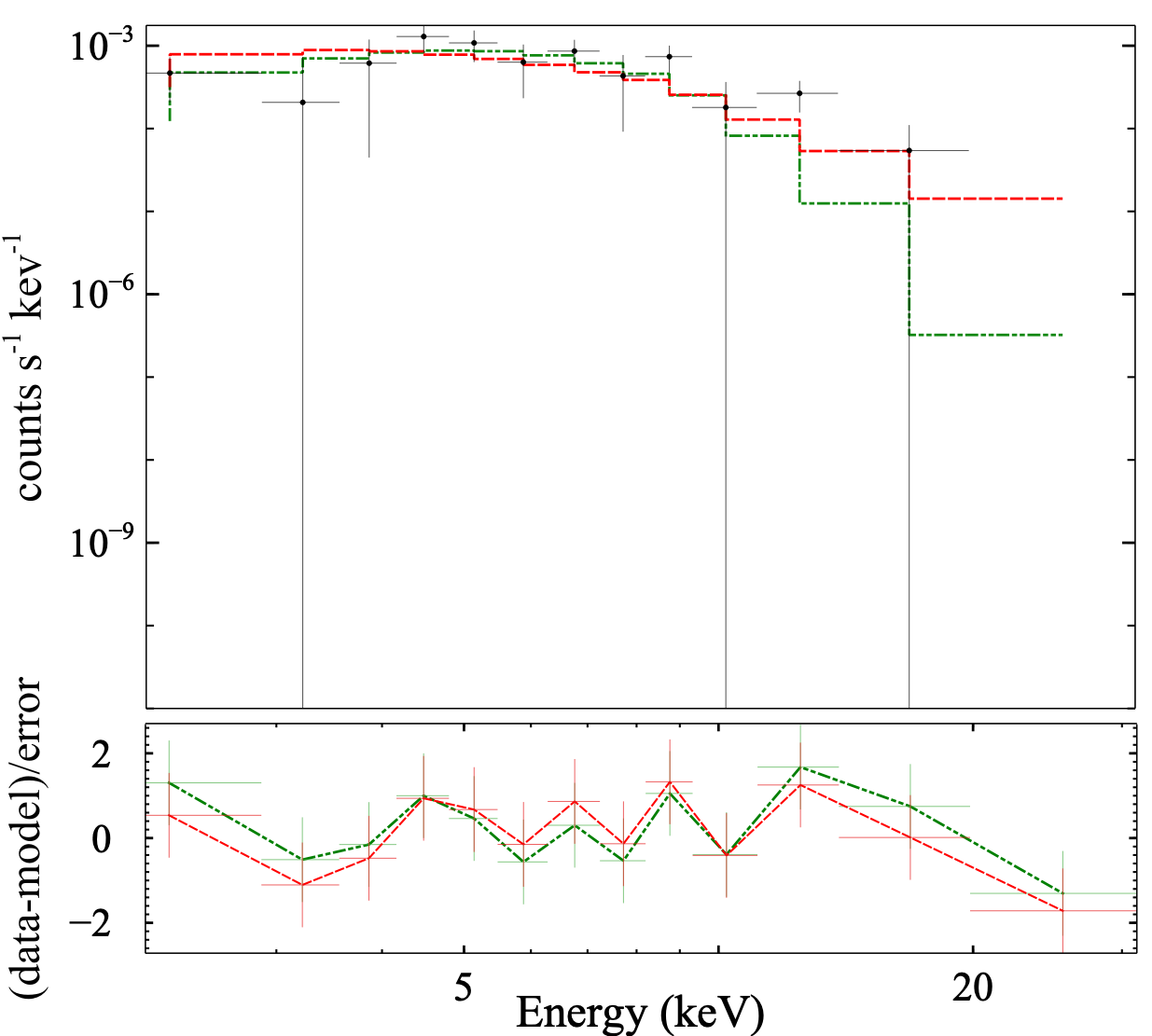}
\end{minipage}
\caption{Broadband energy spectra of the pulsars during Obs I for instruments \textit{NuSTAR}-FPMA (left) and \textit{NuSTAR}-FPMB (right). The energy spectra for both the instruments are fitted with bb*tbabs (green) and po*tbabs (red). The plot below the energy spectra correspond to the spectral residuals for each model.\label{5002}}
\label{fig:test}
\end{figure*}

\begin{figure*}[h!]
\centering
\includegraphics[width=0.49\textwidth]{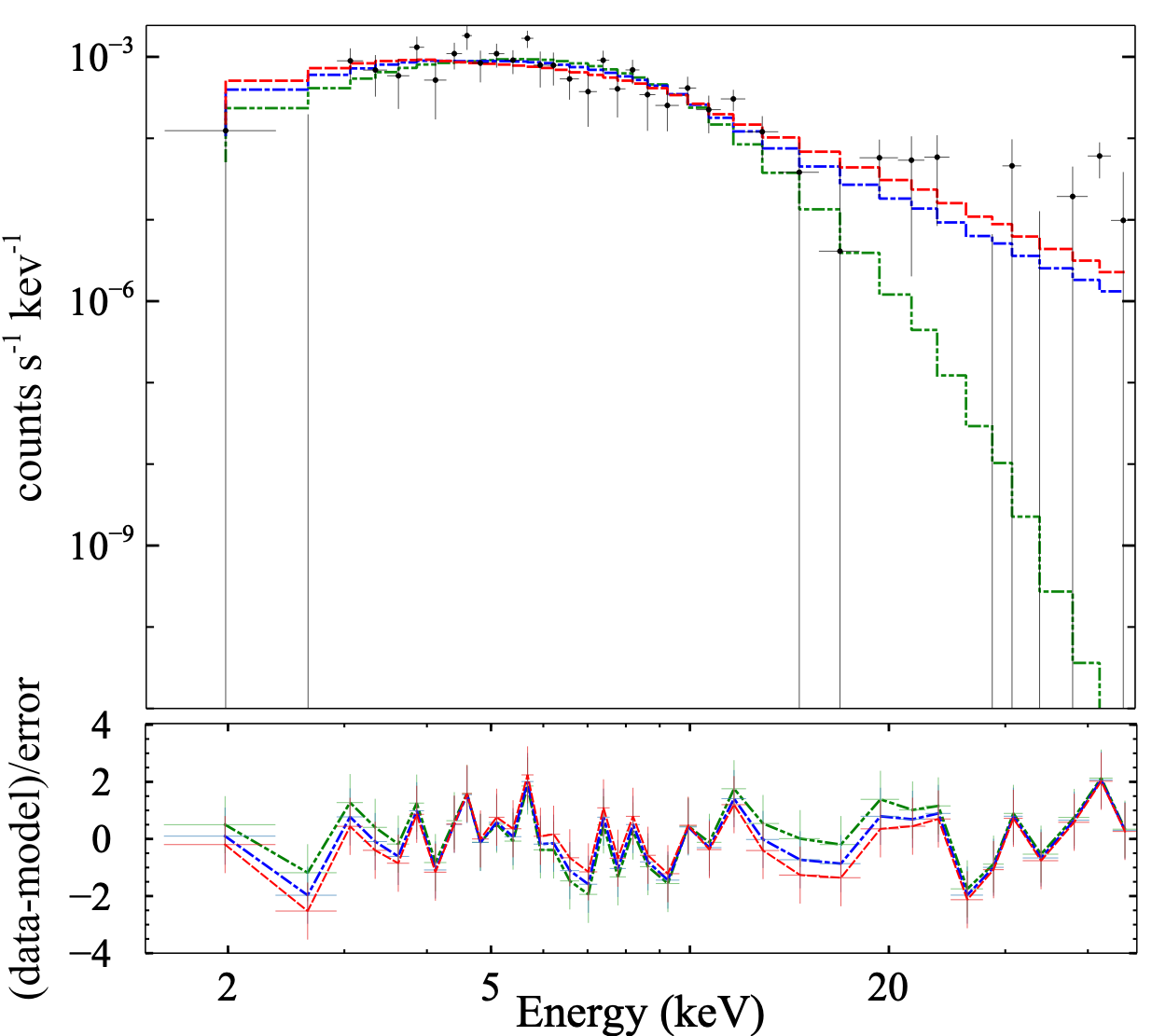}
\includegraphics[width=0.49\textwidth]{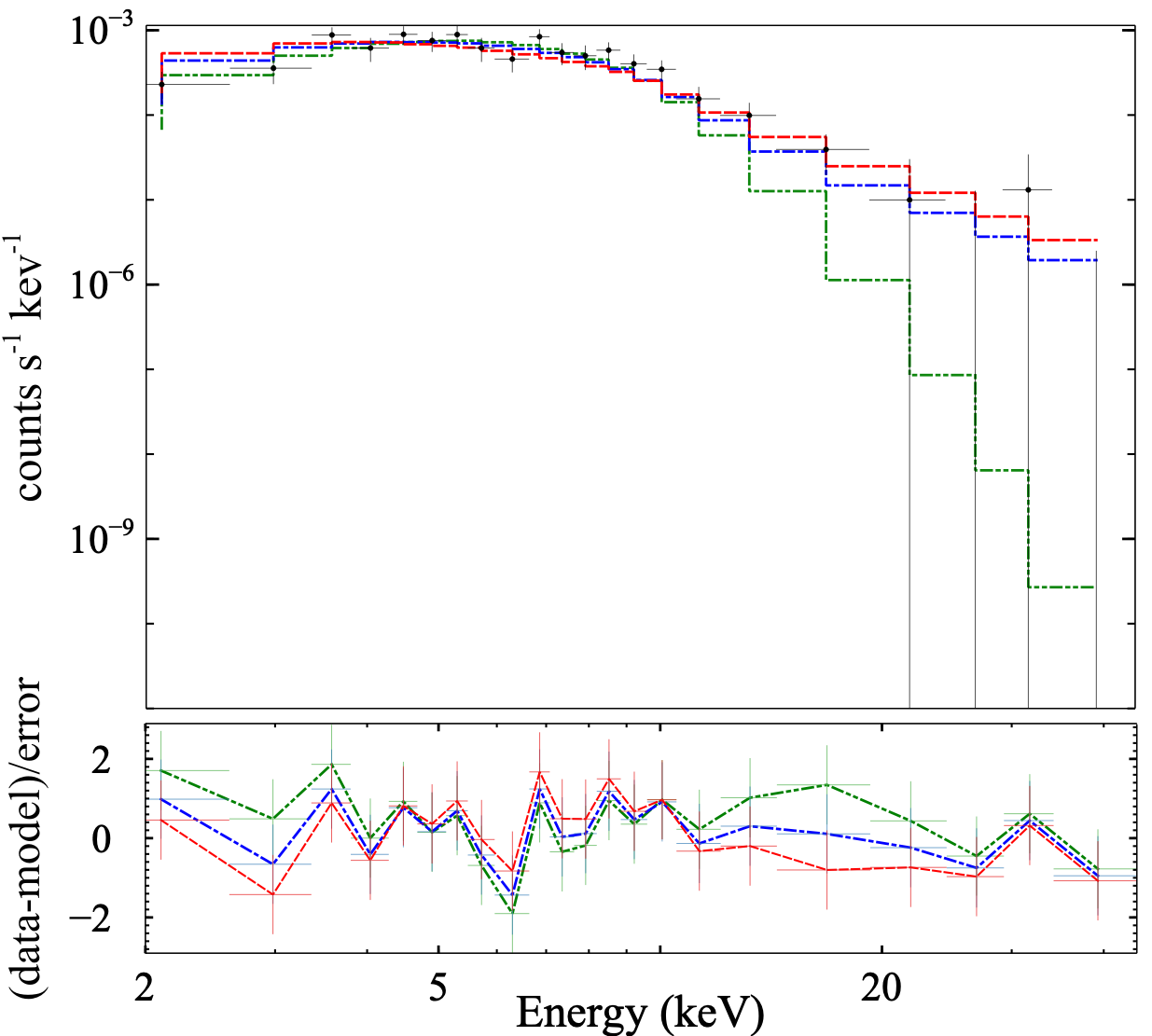}
\caption{Broadband energy spectra of the pulsars during Obs II for instruments \textit{NuSTAR}-FPMA (left) and \textit{NuSTAR}-FPMB (right). The energy spectra for both the instrument is fitted with bb*tbabs (green), po*tbabs (red), and (bb+po)*tbabs (blue). The plot below the energy spectra correspond to the spectral residuals for each model. \label{1002}}
\label{fig:test}
\end{figure*}

\begin{figure*}[h!]
\centering
\begin{minipage}{0.49\textwidth}
    \centering
    \includegraphics[width=1.01\textwidth]{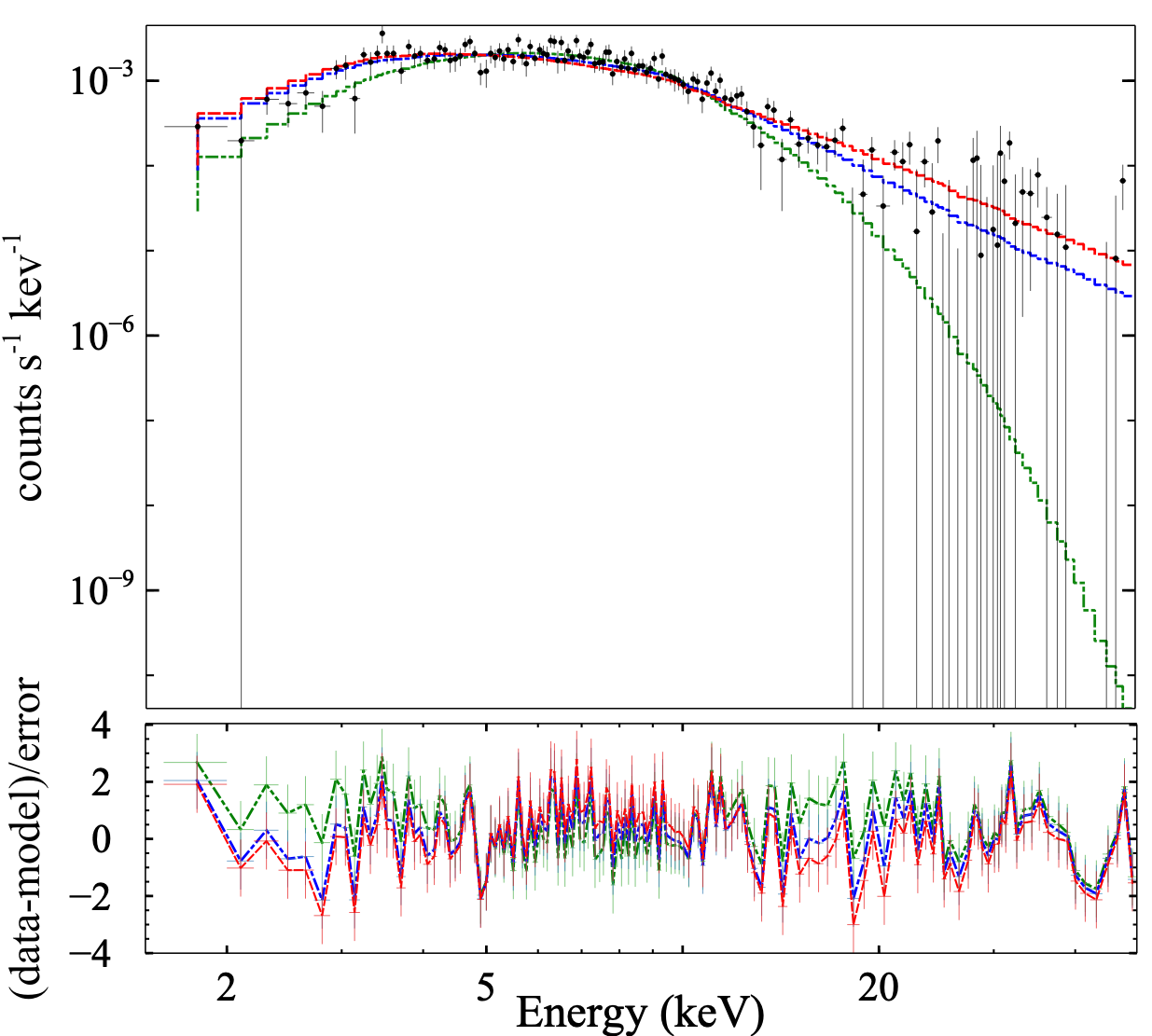}
\end{minipage}
\begin{minipage}{0.49\textwidth}
    \centering
    \includegraphics[width=1.01\textwidth]{fpma_3002_1.png}
\end{minipage}
\caption{Broadband energy spectra of the pulsars during Obs III for instruments \textit{NuSTAR}-FPMA (left) and \textit{NuSTAR}-FPMB (right). The energy spectra for both the instrument is fitted with bb*tbabs (green), po*tbabs (red), and (bb+po)*tbabs (blue). The plot below the energy spectra correspond to the spectral residuals for each model.\label{3002}}
\label{fig:test}
\end{figure*}

\begin{figure*}[h!]
\centering
\begin{minipage}{0.49\textwidth}
    \centering
    \includegraphics[width=1.01\textwidth]{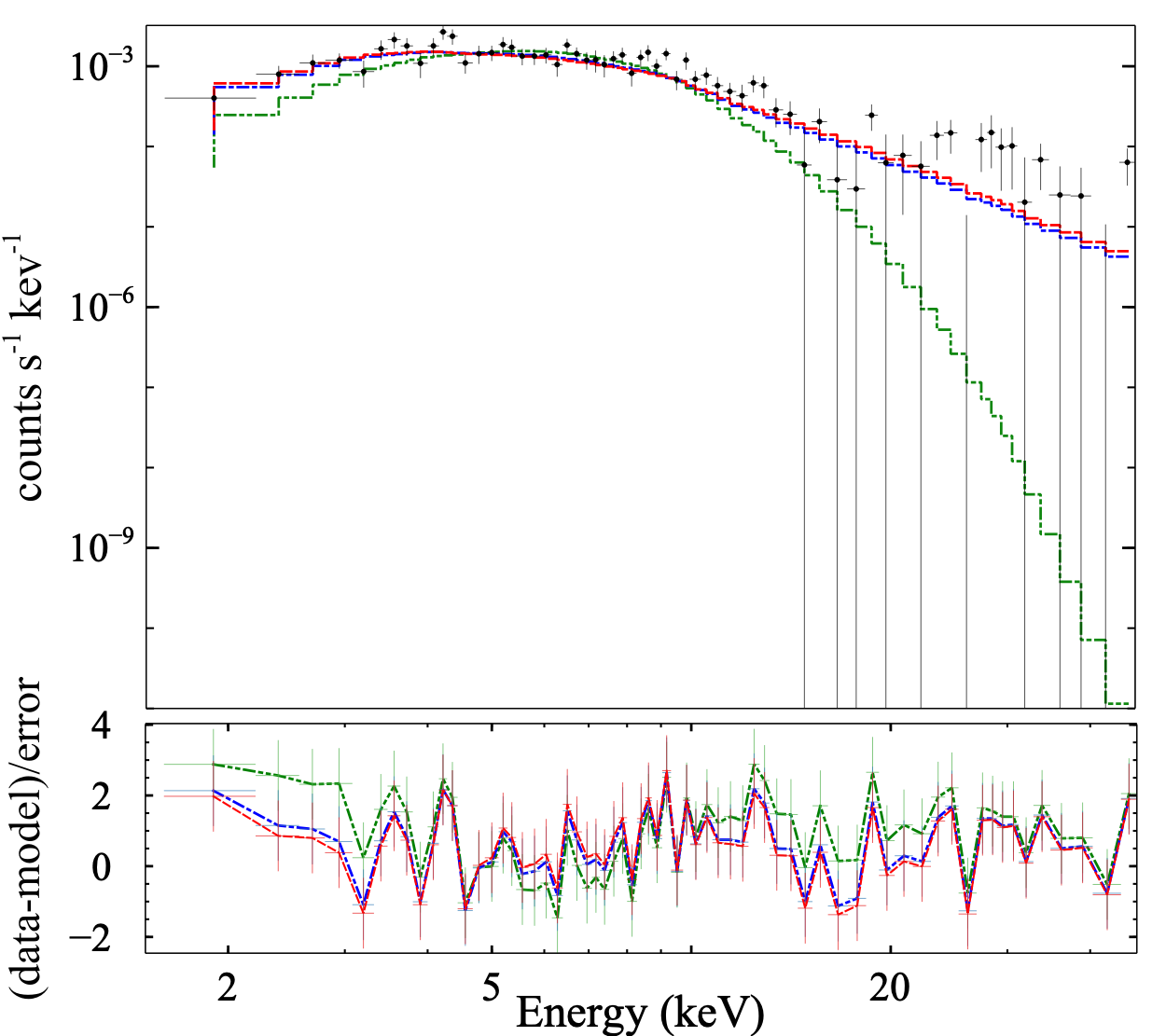}
\end{minipage}
\begin{minipage}{0.49\textwidth}
    \centering
    \includegraphics[width=1.01\textwidth]{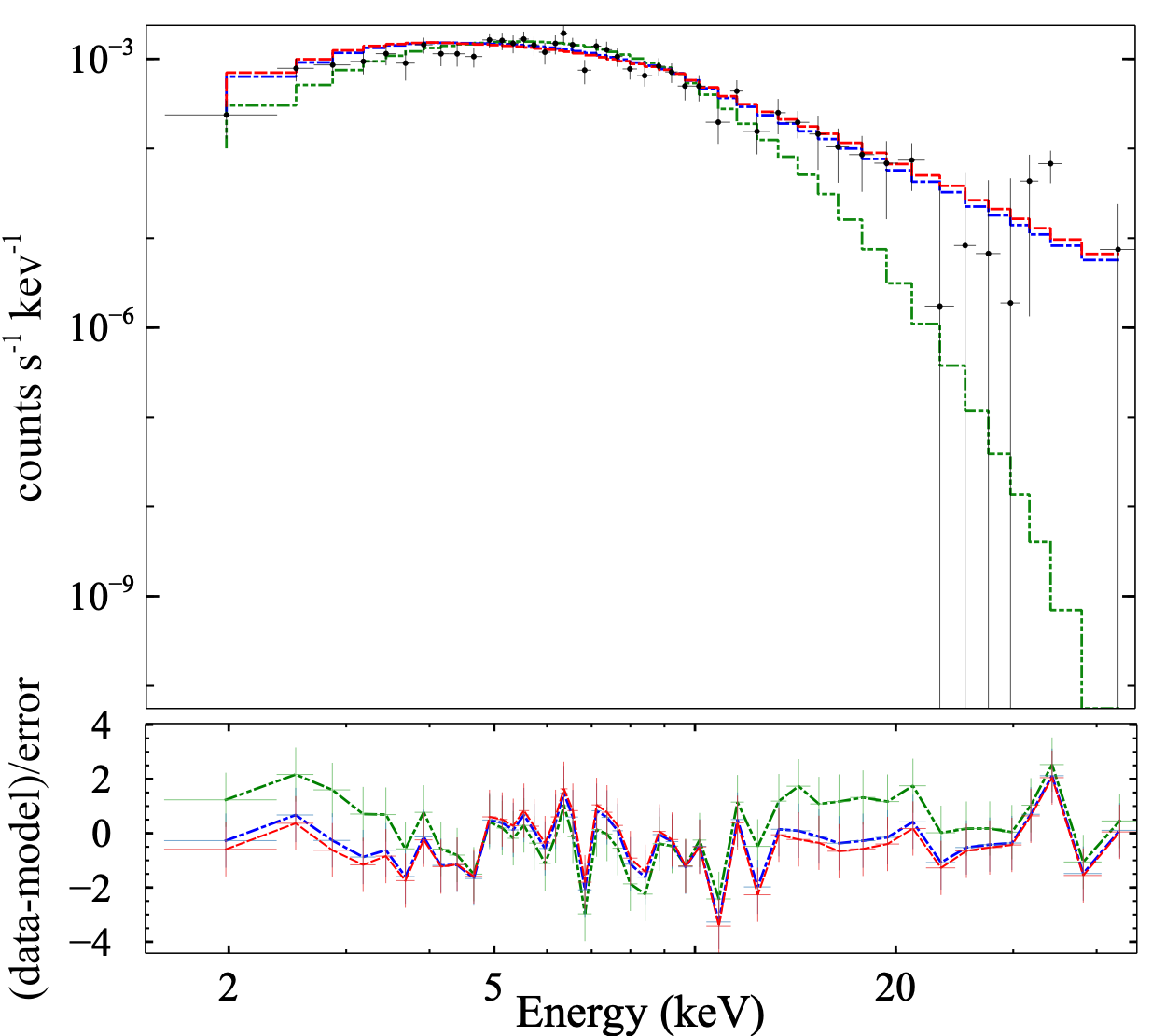}
\end{minipage}
\caption{Broadband energy spectra of the pulsars during Obs IV for instruments \textit{NuSTAR}-FPMA (left) and \textit{NuSTAR}-FPMB (right). The energy spectra for both the instrument is fitted with bb*tbabs (green), po*tbabs (red), and (bb+po)*tbabs (blue). The plot below the energy spectra correspond to the spectral residuals for each model. \label{3004}}
\label{fig:test}
\end{figure*}

The blackbody model provides the best fit for the Obs I energy spectrum. The po*tbabs and (bb+po)*tbabs give similar statistics. However, the single blackbody model is chosen due to high parameter uncertainties of the latter models, especially the power-law index. Figure~\ref{5002} displays the blackbody and powerlaw model fits as well as illustrates low statistics of the observation. Obs II shows the highest null-hypothesis probability for (bb+po)*tbabs but the best $\chi_r^2$ value for po*tbabs. In this case, bb*tbabs show admissible statistics for the fit. Since it is preferable to consider a model where both the models have significant contributions, we use (bb+po)*tbabs to describe the Obs II energy spectrum (Figure~\ref{1002}). For Obs III, (bb+po)*tbabs yields the highest null-hypothesis probability and $\chi_r^2$ closest to 1, indicating the best fit for the data. Although the power-law component remains dominant, Figure~\ref{3002} indicates that its contribution is lower than the previous observation. The bb*tbabs model can be discarded for Obs IV due to its low null hypothesis probability. The po*tbabs and (bb+po)*tbabs models have comparable fitting statistics for Obs IV. The energy spectrum of Obs IV is best described by the (bb+po)*tbabs model, where the contribution of the power law component is dominant, as shown in Figure~\ref{3004}.  

We analyze the best-fit model parameters for each observation to determine the evolution of blackbody temperature and photon index ($\alpha$) over time. Over the first three observations, SXP 138 shows a rapid increase in blackbody temperature from $1.8 \pm 0.2$ keV to $2.5 \pm 0.2$ keV. Although, the temperature in the last observation decreases to $1.8 \pm 0.6$ keV, the uncertainty in the final observation may suggest reduced model fit quality, limiting our ability to draw definitive conclusions about the source's state during this observation. The photon index values might suggest a decreasing trend from $1.8 \pm 0.4$ to $1.7 \pm 0.2$ for Obs II through Obs IV, indicating spectral hardening. We are unable claim this definitively as the three values overlap within their respective error bars.

\begin{table*}[th]
\caption{Summary of spectral fit parameters for different models. Blackbody and Power-law are grouped together, while the other two models are presented in the second and third segment.\label{Para}}\vspace{0.2cm}
\centering
\begin{tabular}{lccc}
\hline
 & \multicolumn{2}{c}{Blackbody}  \\ \hline
Obs & $\rm kT_{\rm BB}$ (keV) & $^{\#}K_{\rm BB} \times 10^6$ & Luminosity$^*$ (${ergs}/{s} $)  \\ \hline
 Obs I$^\dagger$  & $1.8 \pm 0.2$ & $7 \pm 1$ & $2.154\times 10^{35}$ \\ 
 Obs II & $1.9 \pm 0.1$ & $8.0 \pm 0.6$  & $2.516\times 10^{35}$  \\ 
 Obs III  & $2.17 \pm 0.04$ & $12.9 \pm 0.3$  & $4.237\times 10^{35}$ \\  
Obs IV  & $1.92 \pm 0.06$ & $10.0 \pm 0.3$  & $3.176\times 10^{35}$ \\  \hline \vspace{0.3 cm}
\end{tabular}

\begin{tabular}{lccc}
\hline
 & \multicolumn{2}{c}{Power-law} \\ \hline
Obs & $^{\#}\alpha$ & $^{\#}K_{\rm PO} \times 10^4$ & Luminosity$^*$ (${ergs}/{s}$)\\ \hline
Obs I  &  $2.0 \pm 0.3$ & $2 \pm 1$ & $3.857 \times 10^{35}$ \\ 
Obs II & $1.9 \pm 0.1$ & $1.8 \pm 0.4$ & $4.940\times 10^{35}$  \\ 
Obs III &  $1.64 \pm 0.04$ & $1.8 \pm 0.1$ & $9.075 \times 10^{35}$ \\
Obs IV  &  $1.74 \pm 0.06$ & $1.9 \pm 0.2$ & $6.994\times 10^{35}$ \\  \hline \vspace{0.3 cm}
\end{tabular}

\begin{tabular}{lccccc}
\hline
 & \multicolumn{4}{c}{Blackbody + Power-law} \\ \hline
Obs & $\rm kT_{\rm BB}$ (keV) & $K_{\rm BB} \times 10^6$ & $\alpha$ & $K_{\rm PO} \times 10^4$ & Luminosity$^*$ (${ergs}/{s}$)\\ \hline
Obs I  & $2.12 \pm 0.6$ & $5 \pm 6$ & $3 \pm 3$ & $2 \pm 8$ & $2.565 \times 10^{35}$ \\ 
Obs II$^\dagger$ & $1.9 \pm 0.4$ & $4 \pm 1$ & $1.8 \pm 0.4$ & $1.0 \pm 0.8$ & $3.974\times 10^{35}$  \\ 
Obs III$^\dagger$  & $2.5 \pm 0.2$ & $6 \pm 1$ & $1.8 \pm 0.1$ & $1.7 \pm 0.3$ & $6.801 \times 10^{35}$ \\
Obs IV$^\dagger$  & $1.8 \pm 0.6$ & $2 \pm 1$ & $1.7 \pm 0.2$ & $1.5 \pm 0.5$ & $6.445\times 10^{35}$ \\  \hline \vspace{0.3 cm}
\end{tabular}

\footnotesize{$^\dagger$ Represents the model which best fits the observational data (also see table~\ref{Stats} for the fit statistics) \\
$^*$ Calculated using Luminosity = Flux $\times 4\pi d^2$ where \textit{d} is the distance of the SMC from the Earth ($\sim 60kpc$).\\
$^\#$ $K_{\rm BB}$ and $K_{\rm PO}$ represents the norm in the blackbody and power-law models respectively. Photon index is given by $\alpha$.}
\end{table*}


\begin{table*}[t]
\centering
\caption{Comparison of statistical values of the fits for all models across all observations. The best fit is determined by the statistical criteria described in Section~\ref{3.2}\label{Stats}}
\vspace{0.2 cm}
\begin{tabular}{l|cccc|cccc}\hline
\multirow{2}{*}{Observation} & \multicolumn{4}{c|}{bb*tbabs} & \multicolumn{4}{c}{po*tbabs}\\ 
\cline{2-9}
 & $\chi^2$ & d.o.f & $\chi^2_r$ & p & $\chi^2$ & d.o.f & $\chi^2_r$ & p\\ 
\hline
Obs I  & 28.22  & 24  & 1.176 & $2.51\times10^{-1}$ & 29.50  & 24  & 1.229 & $2.02\times10^{-1}$\\
Obs II & 61.25  & 58  & 1.056 & $3.60\times10^{-1}$ & 57.64  & 58  & 0.994 & $4.89\times10^{-1}$\\
Obs II & 466.31 & 304 & 1.534 & $5.81\times10^{-9}$ & 368.88 & 304 & 1.213 & $6.37\times10^{-3}$\\
Obs IV  & 202.52 & 110 & 1.841 & $1.88\times10^{-7}$ & 137.87 & 110 & 1.226 & $3.72\times10^{-2}$ \\
\end{tabular}

\vspace{0.5cm}

\begin{tabular}{l|cccc|cccc}\hline
\multirow{2}{*}{Observation} & \multicolumn{4}{c|}{(bb+po)*tbabs} & \multicolumn{4}{c}{(bb+cutoffpl)*tbabs} \\ 
\cline{2-9}
 & $\chi^2$ & d.o.f & $\chi^2_r$ & p & $\chi^2$ & d.o.f & $\chi^2_r$ & p \\ 
\hline
Obs I  & 27.24  & 22  & 1.238 & $2.02\times10^{-1}$ & 26.23  & 21  & 1.249 & $1.98\times10^{-1}$ \\
Obs II & 50.96  & 56  & 0.910 & $6.65\times10^{-1}$ & 50.38  & 55  & 0.916 & $6.51\times10^{-1}$  \\
Obs III & 335.36 & 302 & 1.110 & $9.05\times10^{-2}$ & 335.60 & 301 & 1.115 & $8.28\times10^{-2}$ \\
Obs IV  & 134.72 & 108 & 1.247 & $4.17\times10^{-2}$ & 135.91 & 110 & 1.235 & $3.10\times10^{-2}$ \\
\end{tabular}
\end{table*}

\section{Discussion \label{4}}
\subsection{Spin Evolution, Accretion Torque and Magnetic field}\label{evol}
The observed spin period variation over time in SXP 138 suggests a long-term spin-down trend. The spin evolution shows a complex profile from Obs. I to Obs. IV. If it's fitted with a linear spin-down trend we get a period derivative of $\dot{P} = 4.27 \times 10^{-9} \, \text{s s}^{-1}$. But the fit is not well constrained, alternatively, we fit the spin periods with two separate linear functions yielding a period derivative of $7.07\times10^{-10}$ss$^{-1}$ and $1.515\times10^{-8}$ss$^{-1}$ between Obs. I and II and Obs II, III, and IV respectively. We also used a quadratic function to measure a possible change in period derivative over time $\ddot{P}$s s$^{-2}$, which measures 4.56$\times$10$^{-16}$s s$^{-2}$. In accreting X-ray pulsars, the spin evolution is primarily governed by the interaction between the neutron star's magnetic field and the accretion disk. The observed spin-down could indicate a continuous decrease in the accretion rate, leading to a reduced angular momentum transfer to the neutron star. It suggests that the source is in a propeller regime\citep{christodoulou2016tracing}, where the magnetospheric radius exceeds the corotation radius, preventing efficient accretion onto the neutron star surface. 

The observed spin-down trend also provides constraints on the neutron star's magnetic field strength (see equation~\ref{bfield}). The spin period evolution is fitted well either by two linear components or a single quadratic component but not a single linear function. Hence, from these variations within a small time, we can assume that the pulsar is in spin equilibrium or near its spin equilibrium, and the equilibrium period $P_{\text{eq}}$ can be estimated from the balance between accretion torque and magnetospheric interaction. From the quadratic fit if we infer an equilibrium period of 142.65 s (see Fig.\ref{figure4}), then the magnetic field will be \citep{davidson1973neutron,alpar1982new}: 
\begin{align}\label{bfield}
    B = 1.8\times10^{13}GR_{NS6}^{-3}\left( \frac{M}{M_{\odot}}\right)\dot{M}_{16}^{1/2}\left( {\frac{P_{rm eq}}{100 s}}\right)^{7/6}
\end{align}
where, $R_{NS6}$ is the radius of the NS in terms of $10^{6}$ cm, M is the mass of NS in solar mass unit and $\dot{M}_{16}$ is the accretion rate in terms of $10^{16}$ g/s. We can assume canonical NS with a mass of 1.4 M$_{\odot}$ and a radius of $10^{6}$cm. The mass accretion rate can be inferred from the source luminosity to be, L = $\eta\dot{M}c^2$. 

If the source is not currently in equilibrium, maybe it is closer to an equilibrium spin period but overall it is showing a long-term spin down. The inferred $\dot{P}$ is consistent with accretion torque models that suggest the matter is not accreted but rather expelled by the centrifugal force of the neutron star, which is also known as the propeller regime. Ultimately this loss of angular momentum causes the neutron star to spin down. 

\subsection{Pulse Profile and Beaming Geometry}
The pulse profile of SXP 138 exhibits a complex structure; in most of the observations, two high peaks with positive and two secondary peaks with negative normalized intensity are observed. The ratio of intensity for these two high peaks varies with both the energy band and observation time. This variability in pulse profile morphology is likely related to changes in the beaming geometry of the emission region (hotspots). Similar changes in beaming geometry have been traced in SMC pulsars using pulse profile modeling before \citep{cappallo2017geometry,cappallo2020geometry,roy2022modeling}. 

Pulse profiles in X-ray pulsars are typically described by the combined effects of relativistic light bending \citep{beloborodov2002gravitational} and the geometrical alignment of the emitting regions (hotpots) on the NS. The emission from the hotspots can consist of both pencil and fan beam components. The observed flux at an inclination angle $i$ and $\theta$ (the angle between the observer line of sight and magnetic dipole axis) is :
\begin{align}
F(\theta, i) = F_{\text{pencil}}(\theta, i) + F_{\text{fan}}(\theta, i),
\end{align}
where the pencil beam component dominates for low accretion rates and is directed along the magnetic axis, while the fan beam emerges due to scattering in the accretion column at higher luminosities. 

The observed beaming can be expressed in this mathematical form using spherical geometric considerations. 
\begin{align}
    \cos(\theta) = \cos i_1 \cos i_2 + \sin i_1 \sin i_2 \cos (2 \pi \phi)
\end{align}
where, $\theta$ is the angle between the magnetic dipole axis and the observer's line of sight, $i_1$ and $i_2$ are the observer's inclination with the rotation axis and angle between the magnetic dipole axis and pulsar's spin axis respectively. $\phi$ is the phase angle, where 0.0 and 0.5 are the face-on and face-off values, whereas 0.25 and 0.75 phases are side views during a complete spin period of the pulsar.   

The observed variations in SXP 138’s pulse profile suggest that the emission is predominately pencil beam from the two hotspots because for all observations the phase differences between two high peaks are $\sim 0.5$, the intensity ratios are further influenced by general relativistic effects such as gravitational light bending, which modifies the visibility and intensity of the secondary pole depending on the compactness parameter of the neutron star \citep{beloborodov2002gravitational}. The variation in the intensity of the two hotspots can either be due to relativistic effects or differences in the area of the emitting hotspots, this degeneracy can be broken using detailed pulse profile modeling. The observed secondary peaks could be related to complex radiative transfer effects within a small (barely formed) accretion column or might suggest minor contributions due to a weak fan beam \citep{cappallo2017geometry}, hence the variations in the observed intensity of the secondary peaks as a function of energy and phase.

The pulse profile shape suggests that the primary emission originates from two dominant hot spots on the neutron star surface, aligned with the observer's line of sight and 180$^\circ$ apart. The presence of secondary peaks suggests an onset of accretion column formation, where variations in scattering and reprocessing also modify the observed emission pattern (fan beam). Changes in the pulse shape over time may also suggest that the beaming pattern is dependent on the accretion rate. The variation over energy bands can be attributed to different emission sites for different processes, with softer (thermal) X-rays arising from a broader region and harder X-rays (non-thermal, comptonized) more collimated along the magnetic field lines.

\subsection{Spectral Characteristics and Accretion Processes}
Spectral analysis of the four observations using a blackbody plus power-law model suggests that both components are required for an optimal fit in most cases. The blackbody temperature ($kT_{\text{BB}}$) and the relative contribution of the power-law component evolve over the course of the observations. In the initial observations, $kT_{\text{BB}}$ is lower, and the power-law component contributes less significantly to the total flux. However, as the accretion rate increases, the blackbody temperature rises, and the power-law contribution strengthens. We also observe the flux continuously increase for the first three observations, whereas the IIIrd and IVth observations show similar flux levels.

This spectral evolution can be attributed to changes in the accretion disk and the associated Compton scattering \citep{becker2005spectral, becker2007thermal}. At lower accretion rates, thermal emission from the neutron star surface dominates, leading to a cooler blackbody temperature. As accretion increases, the inner disk region heats up, increasing the soft X-ray flux. Additionally, increased accretion leads to the formation of an accretion column, where inverse Compton scattering of thermal photons by energetic electrons contributes to the observed power-law component. Although a strong fan beam is not observed, the substructures in the pulse profiles also suggest there can be a weak fan beam from a semi-formed accretion column. This also indicates that the spectral changes (increase in compotization with time) observed in SXP 138 are closely linked to variations in the emission region geometry and accretion rate.

\section{Conclusion \label{5}}
This study, based on a series of \textit{NuSTAR} observations of SXP 138, provides insight into the spin evolution, the nature of the pulse profile, and the spectral characteristics of this Be/X-ray binary. Our analysis highlights the following key results:

\begin{itemize}
    \item The long-term spin-down trend, fitted linearly, gives $\dot{P} = 4.27 \times 10^{-9} \text{s s}^{-1}$. However, a better fit with two linear or a quadratic function suggests a torque change or reversal between observations I and II. Ultimately, the spin-down continues, indicating a propeller regime.
    \item The pulse profile shows two main peaks, suggesting pencil beam emission from asymmetric hotspots, while two secondary peaks indicate complex radiative transfer, possibly a weak fan beam or beaming geometry changes with accretion rate variations.
    \item Spectral analysis requires both blackbody and power-law components. With enhanced accretion, the blackbody temperature rises and the power-law index decreases, likely due to inner disk heating and accretion column formation.
\end{itemize}

These findings provide a comprehensive picture of the accretion and X-ray emission mechanisms in SXP 138, which was not done before in detail. Monitoring SMC pulsars like SXP 138 over long timescales is crucial for understanding their (and overall the population of HMXB pulsars) accretion behavior and spin evolution. Future studies should aim to track long-term spin variations and investigate spectral state transitions to enhance our knowledge of the accretion processes in these systems.

\bibliography{references}{}
\bibliographystyle{elsarticle-harv}

\end{document}